\documentclass[twocolumn]{aastex62}

\newcommand{\gmcc}{g cm$^{-3}$}

\shorttitle{Glitch Behavior ....}
\shortauthors{Basu et al.}
\begin{document}

\title{Glitch Behavior of Pulsars and Contribution from Neutron Star Crust}

%\correspondingauthor{August Muench}
%\email{greg.schwarz@aas.org, gus.muench@aas.org}

\author{Avishek Basu}
\affiliation{National Centre for Radio Astrophysics - Tata Institute of Fundamental Research \\
NCRA-TIFR, Pune University Campus, Ganeshkhind, Pune, Maharashtra 411007, India}

\author{Prasanta Char}
\affiliation{Inter-University Centre for Astronomy and Astrophysics, Post Bag 4, Ganeshkhind, Pune - 411 007, India}

\author{Rana Nandi}
\affiliation{Tata Institute of Fundamental Research - Department of Nuclear and Atomic Physics \\ Homi Bhabha Road, Navy Nagar, Colaba, Mumbai, Maharashtra 400005, India}

\author{Bhal Chandra Joshi}
\affiliation{National Centre for Radio Astrophysics - Tata Institute of Fundamental Research \\
NCRA-TIFR, Pune University Campus, Ganeshkhind, Pune, Maharashtra 411007, India}

\author{Debades Bandyopadhyay}
\affiliation{Astroparticle Physics and Cosmology Division\\Saha Institute of Nuclear Physics, HBNI\\ Sector - 1, Block - AF Bidhan nagar,
Kolkata- 700064 , India}

\begin{abstract}
Pulsars are highly magnetized rotating neutron stars with a very stable rotation speed. Irrespective of their stable rotation rate, many pulsars have been observed with the sudden jump in the rotation rate, which is known as pulsar glitch. The glitch phenomena are considered to be an exhibit of superfluidity of neutron matter inside the neutron star's crustal region. The magnitude of such rapid change in rotation rate relative to their stable rotation frequency can quantify the ratio of the moment of inertia of the crustal region to the total moment of inertia of the star also called as the fractional moment of inertia (FMI). In this paper, we have calculated FMI for different masses of the star using six different representative unified equations of state  constructed under relativistic mean field framework. We have performed an event-wise comparison of FMI obtained from data with that of theoretically calculated values with and without considering the entrainment effect. It is found that larger glitches can not be explained by crustal FMI alone, even without the entrainment.
\end{abstract}


\section{Introduction}

Pulsars are rotating magnetized neutron stars with magnetic field strength of about
$10^{12}$ G. Most pulsars have very stable rotational periods ranging from milliseconds
to a few seconds owing to their compact size and large mass. 
%The period of pulsar slows down due to continuous emission of large amount of radiations
%and particle wind. Irrespective of the fact that the pulsars are very stable rotator
However, some pulsars do show rotational irregularities, such as 
%. One such example is a 
sudden jumps in the spin frequencies, known as pulsar glitches \citep{RadMan1glt}. In the 
last five decades, 529 glitches have been reported in radio pulsars
\citep{lyne1992,shemar.lyne1996,lyne1996,lyne2000,wang2000,kraw2003,Espinoza2011,Yu2013,Fuentes17}.
The reported fractional spin-up during a glitch\footnote{The fractional spin-up is
defined as the ratio of increase in rotation rate $\delta \nu$ to rotation rate $\nu$
at the time of glitch.} range from 10$^{-9}$ to 33 $\times$ 10$^{-6}$ \citep{manhobbs2011}.
Pulsars, such as PSR B0833$-$45, B1046$-$58, B1338$-$62 and B1737$-$30, show glitches 
with fractional spin-up varying over three orders of magnitude. The glitch events can also 
be followed by exponential recoveries of rotation rate   \citep{Yu2013}. It is believed 
that the interior superfluid is responsible for glitches, where the excess angular
momentum of pinned component of superfluid is transferred to the crust at a critical lag
between the differential  rotation of  neutron star's superfluid and non-superfluid 
component \citep{Alpar1985, Alpar1984}. Observations of pulsar glitches and their
recoveries provide important probe of the composition and  structure of neutron stars 
\citep{haskell} in terms of the depth from the surface of the star, where glitches originate.
Glitches can also help in understanding how the superfluid component of the star is coupled 
with the observable crust of the star and whether  the core of the star participates in
the glitch phenomena. We aim to investigate this by comparing theoretical calculations 
with reported glitch measurements in this work.

%Also the rapid cooling of Cas-A can only be explained by superfluidity in the star.\\

One way to achieve this goal is to place constraints on moment of inertia (MoI) of
different parts of the neutron star (NS) participating in the glitch phenomena. 
There exist distinct density regions inside the star, defined by the local density, 
which changes from $\sim$ $10^4$ \gmcc \, to $\sim$ $10^{14}$ \gmcc \, from the surface 
of NS to its center. The outer most layer (outer-crust) consists of 
fully ionized nuclei arranged in BCC lattice structure embedded in degenerate 
electron gas. With increasing density nuclei become more and more neutron-rich 
\citep{BPS, Ruester2006, Nandi2011a}.
% is composed of $Fe$ group 
% elements arranged in BCC lattice structure embedded in degenerate electron gas
% {\bf Rana : cite reference}. 
At the density $\sim$ $10^{11}$ \gmcc, the neutron
drip point is reached and the inner crust begins. The inner crust is composed of 
neutron-rich nuclei arranged in a lattice and immersed in a free electron gas as well as
a gas of dripped neutrons \citep{BBP, Negele1973, Haensel2001, Nandi2011b}, which
are expected to be superfluid \citep{Baldo2005,Sedrakian2006,Chamel2008}.
% with an
% appreciable fraction of the dripped out neutrons in superfluid state
% {\bf Rana/Prasanta : cite refer}.
Beyond the inner crust, the core of the star may
consist of superfluid neutrons, superconducting protons and other exotic matter
\citep{Ginzburg1964, Baldo2005, Sauls1989, Sedrakian2006, Page2006, Chamel2008}. 

Theoretically, MoI of the crust and core components can be estimated by solving the
structure equations with a given Equation of State (EoS).
In the study of pulsar glitches, mostly non-relativistic EoSs are used 
\citep{Andersson2012,Chamel2013,Ho2015,Delsate2016,Li2016,Pizzochero2017}. On the other hand,
EoSs obtained from relativistic mean field (RMF) model that gives a causal description at all 
densities of NS have been widely used in the literature to study various properties of
nuclear matter as well as NS \citep{Glendenning2000, Dutra2014,Oertel2017}. Only \citet{Piekarewicz2014}, employed RMF EoSs for the NS core
to study the glitch phenomena. However, for the inner crust they used a polytropic EoS that
interpolates between the neutron-drip density and the crust-core transition density.
Since the inner crust
%contains the superfluid component and
contributes more than $99\%$ to the crustal MoI,
it is important to have a realistic EoS for the inner crust.
In the literature, the inner crust and the core are often treated separately
and the corresponding EoSs are matched ``by hand" at the crust-core boundary
which is also chosen arbitrarily \citep{Glendenning2000, Read2009, Fattoyev2018}. The precision in this matching processes can
lead to significant differences in the estimation of MoI of different components \citep{Fortin2016}.
Therefore, it is necessary to use an unified EoS where EoSs of both the inner
crust and the core are calculated from same microscopic theory and as a consequence, the crust-core 
boundary is automatically determined. In this article, we calculate MoI
using unified EoSs derived from different variation of RMF model. 
% In the literature, the 
% outer crust, the inner crust and the core are treated separately by matching
% different Equation of State (EOS) at the boundaries {\bf Rana \& Prasanta :
% cite references of past works}. The precision in this matching processes can lead
% to significant differences in estimation of moment of inertia of different components
% {\bf Rana \& Prasanta : cite references of past works}. In contrast, a unified EOS
% calculated in a Relativistic Mean Field (RMF) approach is likely to provide better
% estimates for these quantities {\bf Rana \& Prasanta : cite references of past works
% and add motivation for RMF unified EOS More to be written here as discussed on 
% Apr 19, 2018}. We adopt this approach in our calculations for comparison with 
% observationally inferred MoI.

Pinned neutron superfluid provides an angular momentum reservoir as its rotation 
rate  is determined by the areal vortex density, which is constant as long as it is 
pinned to the crust. At the same time, the crust continuously slows down due to loss
of its angular momentum in particle wind and electromagnetic radiation. At a 
critical lag in this differentially rotating two-component system, superfluid
vortices get unpinned dumping large amount of angular momentum to the crust, which
is seen as a spin up in the crustal rotation rate, usually inferred by timing the
radio pulse \citep{Alpar1985, Alpar1984}. This implies that the fractional spin-up
provides a probe of the extent of angular momentum transfer and hence MoI of the 
crustal pinned superfluid. The ratio of the MoI of crustal pinned superfluid to 
that of the rest of the star, referred to as fractional moment of inertia (FMI), 
can be related to the observed fractional spin-up, allowing a comparison of
theoretical estimates with those from observations \citep{Link1992, Link1999, Eya2017}.
In this context, it is common practice in the literature \citep{Link1999,Andersson2012,Chamel2013,Ho2015,Eya2017}
to define a quantity called activity parameter, which is essentially the average of all glitches observed
in a time window for a given pulsar. This parameter is then used to estimate the FMI for different
pulsars. However, as we are interested in investigating how far the FMI of the crust could explain the observed glitches, it seems 
more appropriate to use individual glitches instead of the average. Based on calculation of FMI in eight individual Vela pulsar glitches\citep{Alpar93}, \citet{Bdatta1993} ruled out one out of 19 EoS for this pulsar. With a better constraint on the maximum mass of NS for different EoS and much larger glitch database 25 years later, this question 
is worthwhile a detailed re-examination. Therefore, in this article, we consider separately each individual glitch cataloged so far \citep{Espinoza2011} to estimate the FMI.

In this paper, we apply a unified treatment of Equation of State (EoS), obtained
in a variety of RMF models, to estimate the fraction of stellar
moment of inertia of the crust and compare it to that inferred from reported 
observations of pulsar glitches. In Section \ref{eos}, the construction of the EoS
is described followed by estimates of relevant MoI using the structure equations
in Section \ref{strt}. We connect these estimates to observables in Section
\ref{observ} and present our results in Section \ref{res}. Discussion of these 
results and conclusions is presented in Section \ref{disc}.

%( The observed radio beam co-rotates with the crust hence any disturbances in the crust is reflected as the irregularity in the time of arrival of the pulse.

%Neutron star being a rotating system makes the inner superfluid to break up into quantized vortices to mimic the macroscopic rotations. Energetics of the systems drives the vortices to pin with the nuclear clusters \cite{Epstein}.The cite of pinning are part of the crust which continuously slows down by loss of energy due to particle wind and electromagnetic radiation, whereas the areal vortex density determines the velocity of the superfluid component, thus relative velocity gradient grows between the superfluid and normal component. When the lag grows to a critical value, 

% Associated with each and every pulsar glitches, one can calculate the ratio of moment of inertia (MoI) of the region of pulsar contributing to glitches to the total MoI of the star. This ratio is called the fractional moment of inertia (FMI), which can be estimated from the spin down rate of the pulsars \cite{Link1992}.

\section{Equation of State}\label{eos}
It has been shown \citep{Fortin2016} that for an unambiguous calculation of NS properties (especially radius and crust thickness) one needs to employ
a unified EoS, i.e., the EoS of crust and core should be obtained within the same many-body theory. 
As the phenomena of glitch are supposed to be very sensitive to the thickness of the crust, we employ only unified EoSs here.
We construct the EoSs of the inner crust and core adopting  the RMF approach, where the interaction between nucleons
is described by the exchange of $\sigma$, $\omega$ and $\rho$ mesons. For our study, we choose EoSs that represent different variation of the RMF model and also
used widely in the literature \citep{Dutra2014}. In particular, we use parameter sets: NL3 \citep{Lalazissis1997} and GM1 \citep{Glendenning2000} that include 
non-linear self-interaction of $\sigma$-mesons, TM1 \citep{Sugahara1994} that has self-interacting terms for both $\sigma$ and $\omega$ mesons,
NL3$\omega\rho$ \citep{Horowitz2001},  which contains self-interaction of $\sigma$ mesons and coupling between $\omega$ and $\rho$ mesons, 
and DDME2 \citep{Lalazissis2005} and BHB$\Lambda\phi$ \citep{Banik2014}, where coupling parameters are considered to be density dependent.
% It has been shown \citep{Fortin2016} that for an unambiguous calculation of NS properties (especially radius and crust thickness) one needs to employ
% an unified EOS, i.e., the EOS of crust and core should be obtained within the same many-body theory. 
% As the phenomena of glitch are supposed to be very sensitive to the thickness of the crust 
%All six EOSs considered here are unified in that sense.
The EoSs of the inner crust along with the crust-core transition density for all parameter sets excluding BHB$\Lambda\phi$ are taken from \citet{Grill2014}, 
whereas the EoSs of core that contains neutrons, protons, electrons and muons are generated by the standard procedure \citep{Glendenning2000,Dutra2014}. 
As EoSs of both the crust and the core are  described by the same parameter set, they can be joined smoothly at the crust-core boundary without any jump
in pressure and density.
For the EoS of outer crust, we choose DH EoS \citep{Haensel2007}. The choice of outer crust does not have any significant 
impact on the observables as the most part of it is determined from the experimentally measured nuclear masses.
The unified BHB$\Lambda\phi$ EoS is obtained following \citet{Banik2014}. Apart from nucleons and leptons, the core EoS of BHB$\Lambda\phi$
also includes $\Lambda$-hyperons, which interact among themselves via the exchange of $\phi$ mesons.
All six EoSs used here give maximum NS masses ($M_{\rm max}$) more than $2M_\odot$ and are therefore compatible with the constraint 
of $M_{\rm max}=2.01\pm0.04M_\odot$ obtained from observation \citep{Antoniadis2013}.
We use these EoSs to estimate the MoI relevant for the present work.

% 
% As the phenomenon of glitch 
% is supposed to be very sensitive to the thickness of the crust, we take EOS of inner crusts from \citet{Grill2014}, calculated using the same parameters
% sets as mentioned above. For the EOS of outer crust, we choose DH EOS \citep{Haensel2007}. The choice of outer crust does not have any significant 
% impact on the observables as the most part of it is determined from the experimentally measured nuclear masses. 
% It has been shown \citep{Fortin2016}
% that for an unambiguous calculation of NS properties (especially radius and crust thickness) one needs to employ an unified EOS, i.e., the EOS of crust
% and core should be obtained within the same many-body theory. {\bf Rana and Prasanta : A prescription for calculating P as a fn of rho in RMF using above
% needs to be described in some more details as we discussed on Apr 19, 2018. What is done ? What are the steps ? What code is used - is it same as published
% before or modified ? How this is different from that in literature (some references) ? Pros and 
% cons of the procedure adopted - We need to highlight this as this is the new part of the paper and this is different from previous works. Please add suitable
% text.}
% The pressure, calculated at a given density, is then used with the structure equations (Section \ref{strt}) to estimate the MoI relevant for this work.

\section{Structure}\label{strt}
The equilibrium structure of a spherically  symmetric, non-rotating NS is calculated from the solutions of Tolman-Oppenheimer-Volkoff (TOV) equations given by,
\begin{equation}
 \frac{dP(r)}{dr}=-\frac{\left[\varepsilon (r) + P(r) \right]\left[M(r) + 4\pi r^3 P(r) \right]}{r\left[r -2M(r)\right]}
  \label{tov1:eps}
\end{equation}
\begin{equation}
\frac{d\nu (r)}{dr} = - \frac{1}{\varepsilon (r) + P(r)} \frac{dP (r)}{dr}
\label{tov2:eps}
\end{equation}
\begin{equation}
\frac{dM(r)}{dr}=4\pi r^{2}{\varepsilon }(r).
    \label{tov3:eps}
\end{equation}
%
%{\bf define P, M, $\nu$ and $\varepsilon$ }
Here, $\varepsilon (r)$, $P(r)$, $M(r)$ and $\nu (r)$ represent the radial profile for energy density, pressure, enclosed mass and the metric potential respectively.
Complimented with an EoS, the structure Equations (\ref{tov1:eps}, \ref{tov2:eps}, \ref{tov3:eps}) are solved numerically to generate the profiles for pressure, 
energy density, enclosed mass etc. and also the total mass and radius of the star for a given value of central energy density.

Assuming the star is rotating uniformly and the angular velocity ($\Omega$ ) is sufficiently slow ($\Omega^2 R^3 << M$) compared to its Kepler limit, the MoI of
a star can be calculated in the slow-rotation approximation  using Hartle's prescription \citep{hartle}. The metric of a slowly, uniformly rotating star can be 
expressed at the first order of spin frequency $\Omega$ as,
\begin{eqnarray}
 ds^2 = -e^{2\nu(r)}dt^2 +e^{2\lambda(r)}dr^2  -2\omega(r)r^2sin^2\theta d\phi dt \\ \nonumber  +       r^2d\theta ^2 + r^2sin^2\theta d\phi^2 ,\end{eqnarray}
where, $\omega$ is the frame-dragging frequency, which satisfies the following differential equation:

\begin{equation} 
\frac{\mathrm{d}}{\mathrm{d}r} \left( r^{4}j(r)\frac{\mathrm{d}\bar{\omega}(r)}{\mathrm{d}r} \right) + 4 r^{3} \frac{\mathrm{d}j(r)}{\mathrm{d}r} \bar{\omega}(r)=0,
\label{eq:omega}
\end{equation}
where $\bar{\omega}(r) = \Omega -\omega (r)$ and $j(r)$ is defined as

\begin{eqnarray} \label{eq:fp3}
j(r)=e^{-(\nu(r)+\lambda(r))}= e^{-\nu(r)} \sqrt{1-2GM(r)/r}, \\ \nonumber ~~~~~~~ \text{for} ~~ r \leq R.
\end{eqnarray}

Now, we can compute the MoI of the star solving Equation (\ref{eq:omega}) augmented with the TOV equations as:

\begin{equation} \label{eq:MoI1}
I_{\rm total}=\frac{8 \pi}{3} \int_0^R dr r^{4} \frac{\left( \mbox{$ \varepsilon $} (r)+P(r) \right)}{\sqrt{1-2GM(r)/r}} \frac{\bar{\omega}(r)}{\Omega} e^{-\nu(r)}.
\label{eq:Itotal}
\end{equation}

To calculate the FMI which is responsible for a glitch, we also calculate the crustal contribution to the MoI separately by performing the integration in Equation (\ref{eq:MoI1}) from the crust-core transition radius to the surface of the star:
\begin{equation} \label{eq:new1}
I_{\rm crust}=\frac{8 \pi}{3} \int_{R_{\rm c}}^{R} dr r^{4} \frac{\left( \mbox{$ \varepsilon $} (r)+P(r) \right)}{\sqrt{1-2GM(r)/r}} \frac{\bar{\omega}(r)}{\Omega} e^{-\nu(r)}
\label{eq:Icrust}
\end{equation}
where $R_{\rm c}$ denotes the crust-core boundary. Note that $I_{\rm crust}$ consists of both the non-superfluid and the pinned superfluid component of the MoI
of crust. Thus, $I_{\rm crust}$ cannot be directly related to FMI using the measurements of observed fractional spin-up during a glitch. We provide a prescription
for comparing this theoretical estimate with observations in the next section.

\section{Interpreting the data}\label{observ}

The pulsed emission in radio waveband is a direct measure of rotation of the crust of the pulsar, which can be very precisely modeled using the pulsar timing technique. 
In pulsar timing, predictions of pulse times-of-arrival (TOAs) from a rotational model of the star are compared with the observed TOAs. The difference between the 
observed and predicted TOAs are called timing residuals, which are minimized in a least-square sense to improve the parameters of the rotational model, thus 
precisely estimating the time evolution of the rotation of the pulsar. The pre and post glitch models shows a sudden jump in the rotation period $\nu$. The fractional
change in $\nu$ is therefore available from observations.

% Along with the $\nu$, in most of the cases post glitch epoch also shows change in the $\dot{\nu}$, the rate of spin down. The relative  change in the spin down rate i.e $\frac{\dot{\nu}_{post-glitch} - \dot{\nu}_{pre-glitch}}{\dot{\nu}_{pre-glitch}}$ $\equiv$ $\frac{\Delta \dot{\nu}}{\dot{\nu}}$ give the lower limit on FMI as given in Eq \ref{gltform}. %ion of the crust gets communicated with the observer via the time of arrival of the pulsed emission. The rotation of the pulsar 
%The model of rotation can very 

%\begin{equation}
%\label{gltform}
%\frac{I_{\rm superfluid}}{I_{\rm total}} \geq \frac{\Delta \dot{\nu} }{\dot{\nu}}
%\end{equation}

%Hence for each of the glitch with non zero relative change in spin period derivative can give the estimate of FMI of that particular glitch event.
%\\Where as the quantity we evaluate is $I_{c}/I$, using Eq \ref{eq:new1} and \ref{eq:MoI1} overestimates the FMI. The factor $\eta$,(with $\eta < 1$) multiplied to $I_{c}/I$ can give a true estimate of FMI. Hence we write 

%\begin{equation}
%\label{etafmi}
%\frac{I_{\rm superfluid}}{I_{\rm total}} = \eta \frac{I_{\rm crust}}{I_{\rm total}}
%\end{equation}
%Here we do not give any prescription for evaluating $\eta$, all our calculations assumes $\eta = 1$, hence the maximum possible FMI.
%\section{calculations}

If we assume that the spin-up is due to the transfer of angular momentum from a superfluid, which is not co-rotating with the crust, then the fractional MoI of this
superfluid to the  MoI of the rest of the star for each glitch is given by \citep{Eya2017}

\begin{equation}
\frac{I_{\rm crsf}}{I_{\rm rest}} = - \frac{1}{\dot{\nu_c}}\frac{\Delta \nu_i}{t_i} 
\end{equation}

where $I_{\rm crsf}$, $I_{\rm rest}$, $\Delta \nu_i$, $\dot{\nu_c}$  and $t_i$ are the MoI of the pinned crustal superfluid (not co-rotating with the crust), 
the MoI of the rest of the star (assumed co-rotating with the crust), the spin-up at the $i^{\rm th}$ glitch, the mean rotational spin-down rate and the time elapsed
before the $i^{\rm th}$ glitch since the preceding glitch, respectively.

The above equation can be simplified as 
\begin{equation}
\frac{I_{\rm crsf}}{I_{\rm rest}} = 2\tau_c \left(\frac{\Delta \nu}{\nu}\right)_i\frac{1}{t_i}
\label{eq:IresbyIc}
\end{equation}

where $\tau_c\, (=-\nu/2\dot{\nu_c}$) is the characteristic age of the pulsar. Since $t_i$ is the time preceding the last glitch, one cannot calculate the $I_{\rm crsf}/I_{\rm rest} $ 
for the first glitch.
We want to connect this $I_{\rm crsf}/I_{\rm rest}$ with $I_{\rm crust}/I_{\rm total}$.
Assuming no contribution in the angular momentum transfer from the core superfluid  we have, 
\begin{eqnarray*}
\frac{I_{\rm crust}}{I_{\rm total}} &=& \frac{I_{\rm crsf} + I_{\rm crnsf}}{I_{\rm crsf}+I_{\rm rest}}\\
\frac{I_{\rm crust}}{I_{\rm total}} &>& \frac{1}{1+I_{\rm rest}/I_{\rm crsf}}
\end{eqnarray*}
where, $I_{\rm crnsf}$ is the non-superfluid (and hence co-rotating with the crust) component of crustal MoI. Now, 
since $I_{\rm rest}>>I_{\rm crsf}$, we have $I_{\rm crust}/I_{\rm total}> I_{\rm crsf}/I_{\rm rest}$ and we can write using Eq. (\ref{eq:IresbyIc}):
\begin{equation} \label{eq:FMIobs}
\frac{I_{\rm crust}}{I_{\rm total}}>2\tau_c\frac{1}{t_i}\left(\frac{\Delta\nu}{\nu}\right)_i,
\end{equation}
This connects our theoretical estimate of FMI with that inferred from observations and this equation has been used in the next section to produce the histograms estimating the
$I_{\rm crust}/I_{\rm total}$. 

When the entrainment coupling between the neutron superfluid and the crustal non-superfluid component is taken into consideration, we get \citep{Andersson2012}
\begin{equation}
\frac{I_{\rm crust}}{I_{\rm total}}>2\tau_c\frac{<m_n^*>}{m_n}\frac{1}{t_i}\left(\frac{\Delta\nu}{\nu}\right)_i,
\end{equation}
where $<m_n^*>\over m_n$ is the ratio of average effective mass of neutrons in the inner crust and of bare neutron mass and has value in the range $4.3-5.1$ due to the entrainment effect \citep{Chamel2012, Andersson2012,Delsate2016}.

\section{Results}\label{res} 
\begin{figure}
\centering
\includegraphics[scale=0.25]{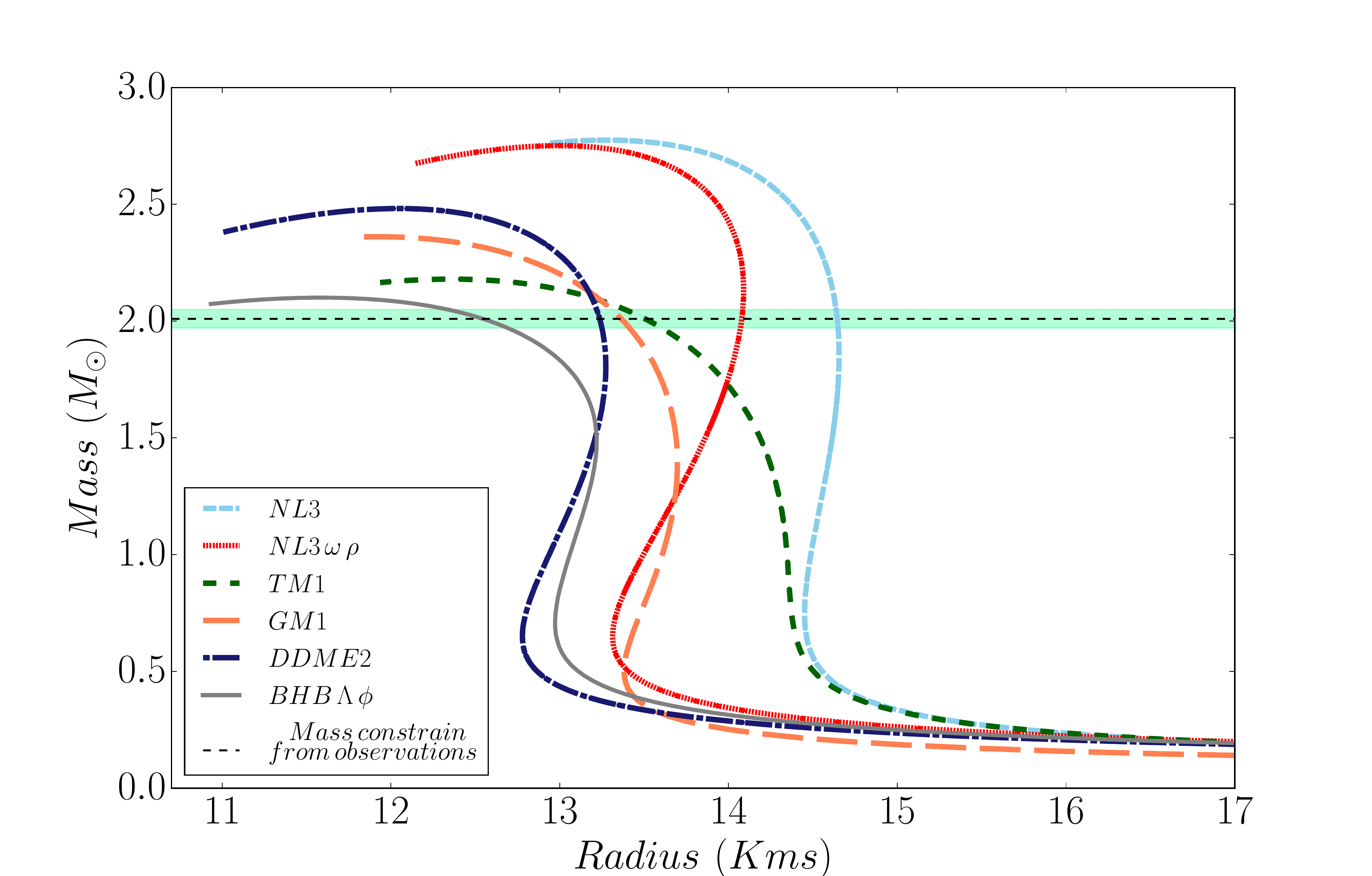}
\caption{The mass radius diagram for NL3 (sky blue), NL3$\omega \rho$ (red), TM1 (green), GM1 (orange), DDME2 (dark blue) and BHB$\Lambda \phi$ (grey) EoS. The blue dotted line represents the mass constrain of 2.01 $\pm$ 0.04 $M_{\odot}$ from observation \citep{Antoniadis2013}. }
\label{fig:MRdia}
\end{figure}
We have used six different unified equations of state NL3, NL3$\omega \rho$, GM1, TM1, DDME2 and BHB$\Lambda\phi$ as described in Section \ref{eos}. 
The mass radius sequences of these six EoSs are shown in Figure \ref{fig:MRdia}. All these EoSs satisfy the present observational constraint on the neutron star
maximum mass (2.01$\pm$ 0.04) $M_{\odot}$ \citep{Antoniadis2013}. It is to be noted that color code to represent various EoSs in Figure \ref{fig:MRdia} has been uniformly applied in all the diagrams throughout the paper. 

\begin{figure*}
%\begin{tabular}{cc}
\includegraphics[scale=0.27]{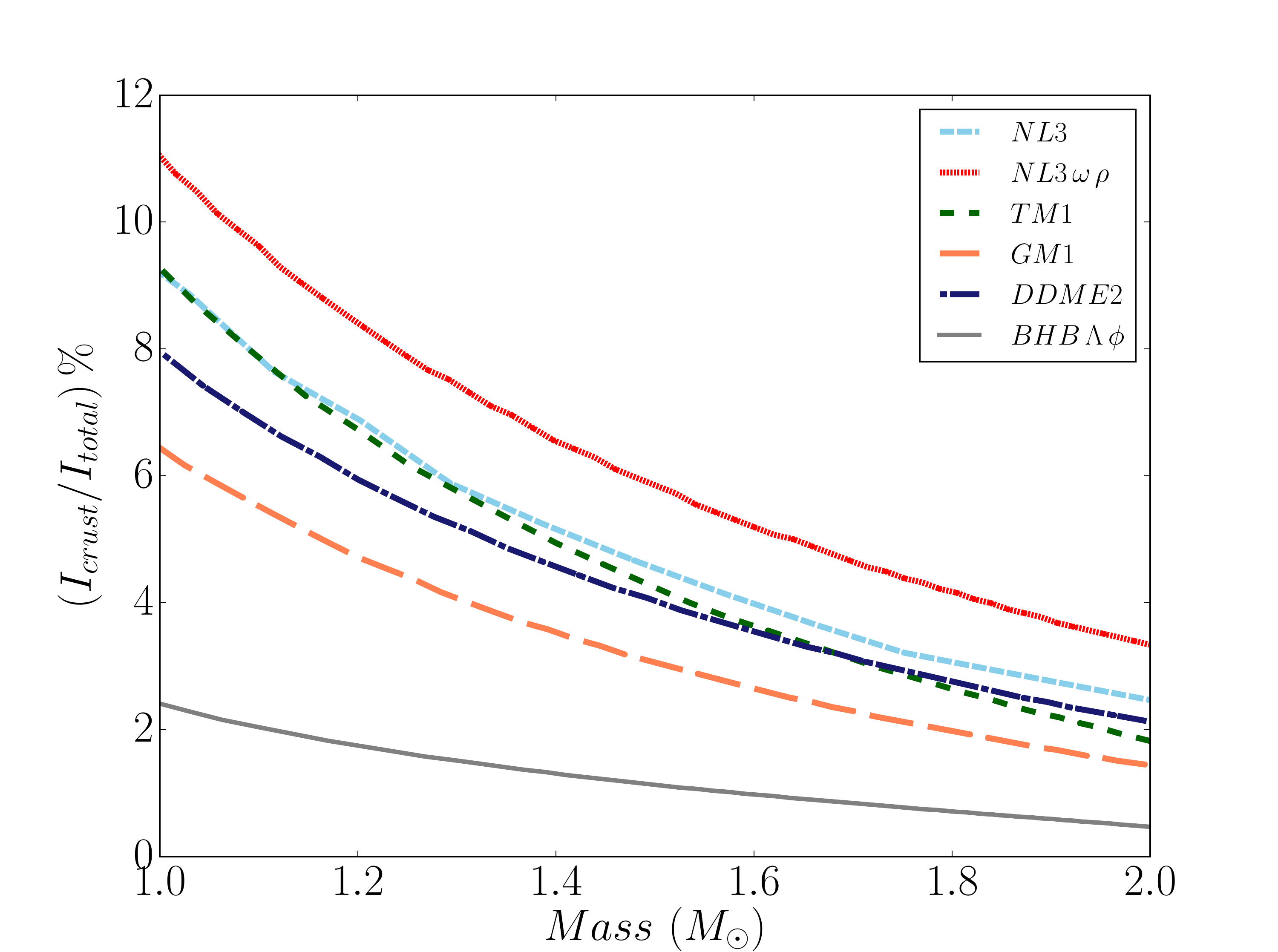} 
\includegraphics[scale=0.27]{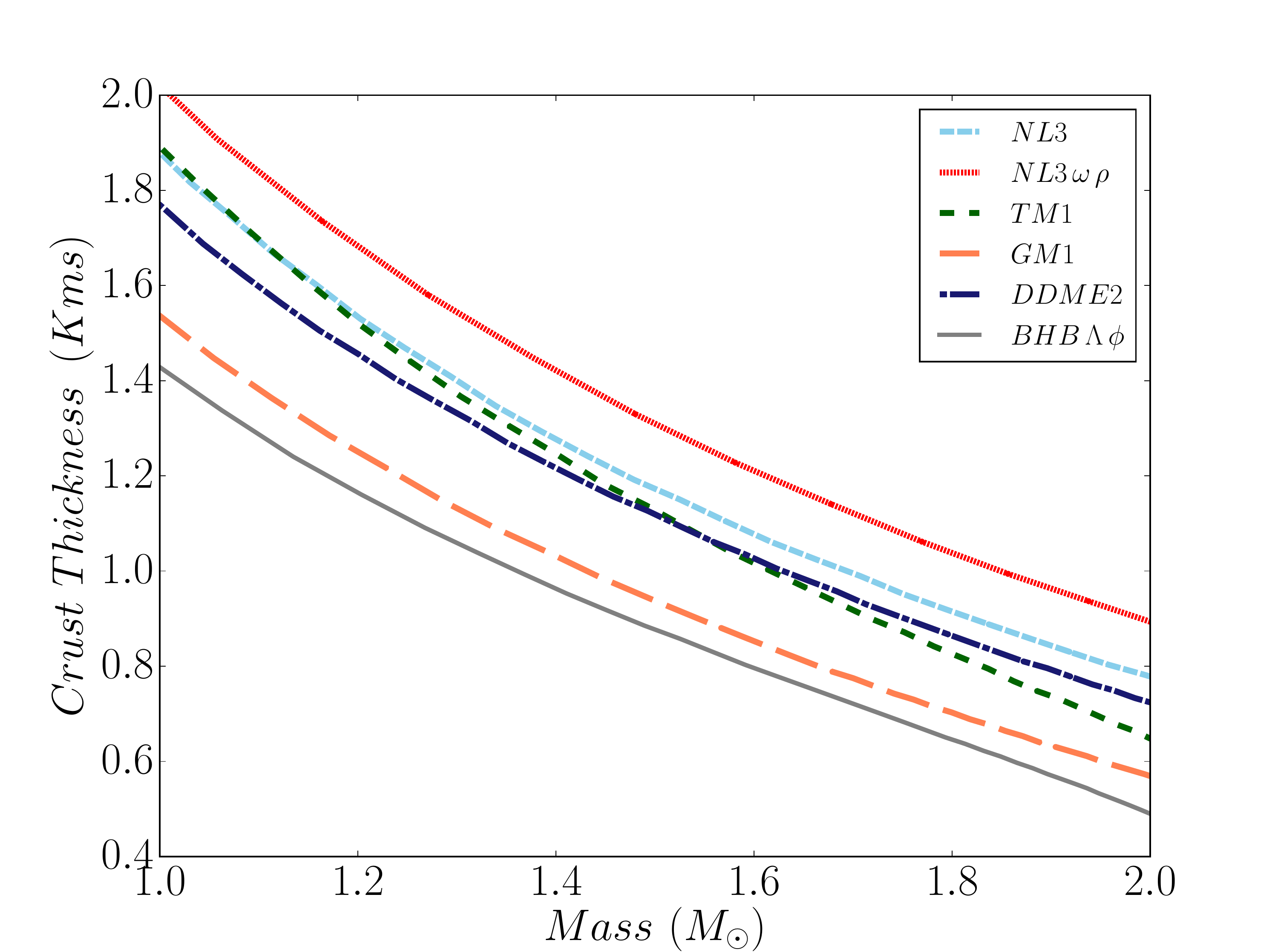} \\
%\end{tabular}
\caption{(\textit{Left}) The fraction of MoI of the crust as compared to the total MoI (expressed in percentage) for six different EoSs as a function of stellar mass.
(\textit{Right}) The crust thickness in kilometers as a function of stellar mass.}
\label{fig:crust}
\end{figure*}
We have calculated the fraction of MoI of the crust to the total MoI as a function of mass of the stars in percent for each of the six EoSs using Equations (\ref{eq:Itotal}) and (\ref{eq:Icrust}) and the results are shown as percentage  in the left panel of Figure \ref{fig:crust}. 
The distance of the crust-core boundary from the center of the star ($R_c$) as well as the radius of the star for each EoS are obtained from
the solution of TOV equation. Hence, one can readily calculate the crust thickness as $\Delta R = R - R_{\rm c}$. The right panel of Figure \ref{fig:crust}
shows the crust thickness  as a function of the stellar mass. It can be clearly seen that the thickness of the crust can be very large $\sim (1.4-1.8)$ km for a lighter
star of mass $\sim 1 \, M_{\odot}$ whereas, for massive stars of mass $\sim$ 2 $M_{\odot}$ it can be as small as $\sim0.5$ km.
Observing the Figure \ref{fig:crust}, it is very much evident that the relation of the crust thickness and its FMI with the stellar mass go hand in hand.
Instead of unified EoS if a polytropic EoS is used for the inner crust as done by \citet{Piekarewicz2014}, we find that depending on the choice of the polytropic index the value of FMI gets overestimated by $0-8\%$ for a $1.4M_\odot$ NS for NL3$\omega\rho$ EoS and the error is more for low mass stars. Similar results are expected for other EoSs as well.
% {\bf If a polytropic EoS of form $P = A + B\epsilon^{\gamma}$ is used for inner crust the values of FMI get overestimated for given mass and $\gamma$. For example, depending on the choice of $\gamma$ the error in the estimation of FMI can be $0-8\%$ for a $1.4M_\odot$ for NL3$\omega\rho$ EOS and the trend increases with lower mass. Similar result is also expected for other EoS.  }

\begin{figure*}
\begin{center}
\includegraphics[scale=0.45]{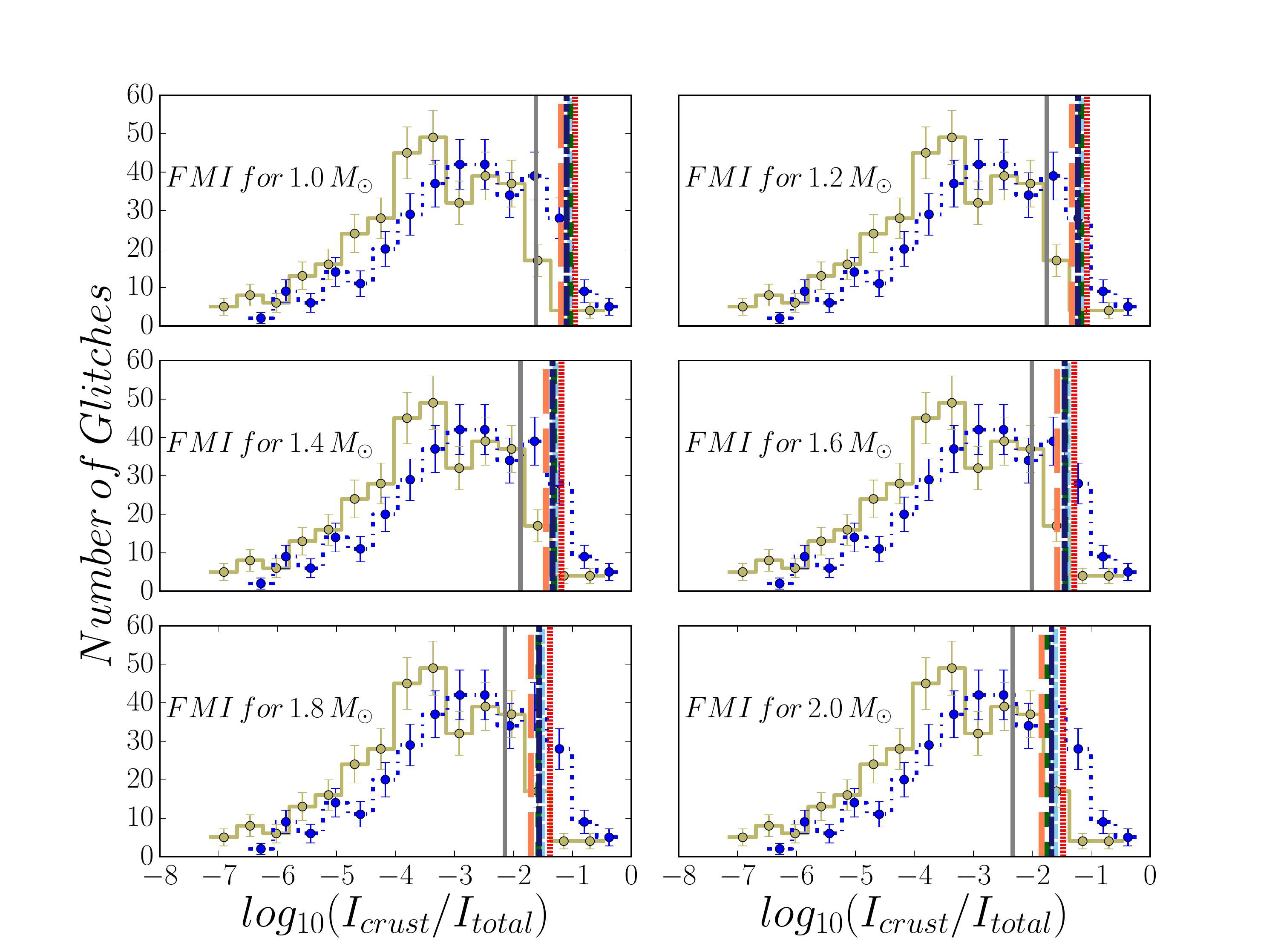}
\caption{Figure shows the distribution of $I_{crust}/I_{total}$. The vertical lines in each plot correspond to the fractional moment of inertia for six 
different equations of states. The plots from top left to bottom right correspond to mass of 1, 1.2, 1.4, 1.6, 1.8 and 2.0 $M_{\odot}$ respectively. The distribution plotted with yellow line corresponds to the distribution without entrainment, whereas the distribution plotted with blue dotted line has been constructed taking entrainment value $\frac{<m_n^*>}{m_n} \, = $ 4.35 into account. Both distribution are obtained from 335 glitches.}
\end{center}
\label{fig:td}
\end{figure*}
%In figure 4,5,6,7,8 we have constructed the distribution of $log_{10}(\frac{\Delta \dot{\nu}}{\dot{\nu}})$ 
%from the publicly available data on pulsar glitches at Jodrell bank website \footnote{http://www.jb.man.ac.uk/~pulsar/glitches/gTable.html} \cite{Espinoza2011}. 
The values of FMI estimated from all the observed glitches  cataloged so far \citep{Espinoza2011}
%maintained by Jodrell Bank Centre of Astrophysics
%\footnote{See the glitch data (as of December 2017) at http://www.jb.man.ac.uk/pulsar/glitches.html \citep{espinoza}}
are shown in Figure \ref{fig:td}, where the distribution of $log_{10}(\frac{I_{\rm crust}}{I_{\rm total}})$, estimated using eq. (\ref{eq:FMIobs}) for all
observed glitches, are plotted for different assumed masses of the NS (1.0, 1.2, 1.4, 1.6, 1.8 and 2.0 $M_{\odot}$ respectively). The error bars given on each 
bins are Poissonian error bars. The vertical lines are constraints from our theoretical calculations of $log_{10}(\frac{I_{\rm crust}}{I_{\rm total}})$ for six 
EoSs considered in this work.  We note that the observed $I_{\rm crust}/I_{\rm total}$ always overestimates the superfluid reservoir as it assumes the crust is made
entirely of pinned superfluid. Hence, the actual contribution from the superfluid in the crust towards glitch is smaller than the estimated
$I_{\rm crust}/I_{\rm total}$. In all the cases, the NL3$\omega \rho$ model can explain a larger FMI as it has the largest crust (Fig. \ref{fig:crust}) . It is evident that there is a significant fraction of the glitch events,
which cannot be explained solely by the crustal superfluidity. The number of such events (out of 335 glitches), which cannot be  explained by the crustal
superfluidity alone for all six EoSs and six different assumed NS masses are tabulated in Table \ref{tab:glitch}. The estimate of $log_{10}(\frac{I_{\rm crust}}{I_{\rm total}})$
considering entrainment are also plotted in the same figure. As expected, the fraction of glitches, which cannot be explained entirely by the crustal superfluidity
alone is larger in presence of entrainment.\\

From the Table \ref{tab:glitch}, we can clearly conclude that at least 1 to 20 \% glitch events without entrainment require the angular momentum reservoir of pinned superfluid 
located outside the crust, even in the limiting case of the crust entirely made up of pinned superfluid. When the entrainment is taken into account the percentage changes to 4 \% to 36 \% .Therefore, we must consider the possibility of the 
contribution of the core in these glitch events. 
%The upper and the lower limits of FMI are given by our calculation of $\frac{I_{c}}{I}$ and from the prescription in \cite{Link1992} Eq \ref{gltform} respectively.
\begin{table*}
\centering
\begin{tabular}{|l|l|l|l|l|l|l|l|}
\hline
*& 1.0 $M_{\odot}$& 1.2 $M_{\odot}$& 1.4 $M_{\odot}$& 1.6 $M_{\odot}$& 1.8 $M_{\odot}$& 2.0 $M_{\odot}$\\
\hline
NL3& 4(16)& 5(22)& 8(31)& 8(39)& 11(71)& 13(60)\\
\hline
NL3$\omega \rho$&4(12)& 5(17)& 5(25)& 8(31)& 8(37)& 11(47)\\
\hline
DDME2&5(17)& 7(28)& 8(34)& 10(43)& 13(56)& 16(66)\\
\hline
TM1&4(16)& 5(23)& 8(32)& 10(43)& 13(56)& 18(70)\\
\hline
GM1&5(25)& 8(33)& 10(43)& 13(56)& 17(68)& 28(80)\\
\hline
BHB$\Lambda \phi$&14(61)& 21(71)& 31(81)& 37(94)&53(106)&68(121)\\
\hline
\end{tabular}
\\ 
\caption{Table shows the number of glitch events that cannot be explained from crustal superfluid alone. The quantities in parenthesis shows glitch event when entrainment ($\frac{<m_*>}{m} \, = 4.35$) is taken into account for six different assumed NS masses (columns) and six different EoSs (rows).}
\label{tab:glitch}
\end{table*}

\section{Conclusions and Discussions}\label{disc} 

In this work, a unified treatment of EoS, using a relativistic mean field  approach, has been carried out to estimate the fraction of MoI of the crust compared to the total NS MoI. 
We have used individual glitches instead of the average of glitches defined via activity parameter, in our estimation
of FMI.
We compared the theoretical estimates with those obtained from observations and conclude that about a few percentage of glitches cannot be explained by angular  momentum transfer from the pinned superfluid in the crust alone, even without the entrainment effect. The fraction of such glitches range from 1 to 20 \% for different EoSs and assumed NS masses. 
%This fraction is larger if entrainment is included as expected.

In this context, it is important to note that the recent binary neutron star merger event GW170817 has provided unique insights on the global properties of isolated neutron stars \citep{Abbott2017}. Several authors have shown that GW170817 sets an upper limit of $\sim 2.1-2.2M_\odot$ on the maximum mass of a NS \citep{margalit2017,shibata2017,rezzolla18,ruiz2018}. The radius of a $1.4M_\odot$ star has also been constrained to $\lesssim$ 13.5 km \citep{most18,Abbott2018,Nandi2018}. In this paper, we have used both stiff EoS (NL3) and moderately soft EoS (BHB$\Lambda \phi$) for representative purpose.  
%In view of those studies, one may ask the necessity of using a very stiff EoS in our studies of FMI and glitches. 
It is evident from Figs \ref{fig:MRdia} and \ref{fig:crust} that  stiffer EoS, like NL3, usually leads to a larger radius. This makes the crustal thickness as well as the $I_{\rm crust}/I_{\rm total}$ larger than that of a moderately softer EoS.   While a majority of glitches can be explained by transfer of angular momentum from crustal superfluid for such stiffer EoS,   some glitches still remain unexplained. On the other hand, a larger number of glitches are inconsistent with the transfer of angular momentum solely from crustal superfluid for a moderately softer EoS, such as BHB$\Lambda \phi$, which predict a maximum mass and radius consistent with the upper limit from GW170817 unlike stiffer EoS \citep{Bhat2018}.  Therefore, the conclusion  that only crustal contribution to the angular momentum transfer can not  account for all the glitches becomes stronger   if   the constraints from GW170817 are taken into account.

The number of unexplained glitches is even larger if entrainment effect is taken into consideration, as expected.  It is worth mentioning here that a
 recent study \citep{Watanabe2017} showed that the effect of entrainment is not that significant, as estimated earlier \citep{Chamel2012}.
 Thus, our calculations suggests that the superfluid in the core is also likely to participate at least in the larger glitches.
The rotation of crustal superfluid is believed to be constrained due to pinning of vortex to the nuclei in the crustal lattice, thus conserving the areal density of vortices \citep{Sauls1989}.  In order for the superfluid in other parts of the star, particularly the core, to participate in a glitch, similar constraint is required on this fraction of superfluid, which was generally believed to co-rotate with the crust and slows down in synchronism by expelling vortices. Our analysis also confirms the presence of extra angular momentum reservoir along with the NS crust.
%This requires a constraining force and one possibility is the magnetic field.

For a glitch to happen in the star, it is essential to pin superfluid vortices to some structures, which can co-rotate with the stellar crusts. One such possibility is presence of mixed state at certain depth of the stellar core. The region is marked by presence of confined (hadronic) and de-confined (quark) matter co-existing in equilibrium. The energetics of such region forces the non-dominating component to form crystal structures of various shapes evolving with the density. They could be a potential zone of extra angular momentum reservoir \citep{Glendenning2000}. Similar kind of crystalline structure can also form due to a mixed phase of kaon condensates inside nuclear matter as a first order phase transition \citep{Glendenning1999}. 
The other possible way of pinning could be between Abrikosov fluxiods along the magnetic moment due to presence of paired proton superconductor \citep{Sauls1989, WCGHO} at the core of the NS  with Onsagar-Feynman vortices along the rotation axis \citep{dipankar}. This novel mechanism of inter-pinning between fluxiods and vortices were used in literature to expel the magnetic fluxoids from core to the crust of star via the spinning down of the NS. But this mechanism may not probably help in building up of angular momentum, which can be expelled at once to produce a sudden jump in angular velocity as observed in   glitches. 

Recently, \citet{guge14} have suggested pinning of superfluid in outer core by a toroidal field similar to pinning of crustal superfluid with a FMI of superfluid associated with this toroidal magnetic field (I$_{tor}$/I$_{total}$) of the order of 0.3$-$1.2 $\times$ 10$^{-2}$\citep{guge17}. This presents an attractive alternative for contribution from superfluid in outer core mediated by magnetic field as a source of extra MoI reservoir for glitches not explained by crustal superfluid. Large 
glitches in PSR B2334+61 and J1718-3718 could be explained by this mechanism \citep{guge16}. It may be noted that a small fraction of core MoI is needed to explain larger FMIs in our sample as expected, considering the fraction of  core superfluid associated with this toroidal magnetic field to be  much smaller than total MoI of core. Thus, this mechanism can 
potentially explain the glitches in our study, where the crustal superfluid is not enough.

Interestingly, in pulsars, such as PSR B0833$-$45, B1046$-$58, B1338$-$62 and B1737$-$30, we see both small and large glitches. The glitch size varies by a factor of 258, 160, 236 and 3811 in these pulsars respectively. Our calculations suggest that
in all these pulsars atleast one glitch event cannot be explained by BHB$\Lambda \phi$ EoS, whereas other EoSs used can explain the glitches from the crustal angular momentum reservoir except for PSR B1046$-$58. In case of PSR B1046$-$58 the glitch which had $I_{\rm crust}/I_{\rm total} \, = \, 33 \% $, cannot be explained by any of the EoSs used taking even $1.0 \, M_{\odot}$ as the fiducial mass of the NS.
%{\bf nn,mm,oo,pp} glitches respectively in these pulsars cannot be caused by angular momentum transfer from the pinned crustal superfluid (in the limit that the crust is entirely made of superfluid).
Thus, we conclude at least some of the glitches require a participation of core, whereas the smaller glitches can be caused by the crust alone. This also may provide a probe for the strength of a coupling agent and the angular momentum transfer mechanism from core to the crust of the star. Future theoretical calculations to probe this aspect are motivated from this work .
%This also may provide a probe for the strength of a coupling agent, such as the magnetic field just as the glitch size provides an upper limit on the pinning force for the crustal glitches{\bf Avishek do a literature review and cite ref}. Future theoretical calculations to probe this aspect are motivated.

\section*{Acknowledgements}
We would like to thank H. Pais for kindly providing us the EoS table for the GM1 inner crust. P. Char acknowledges support from the Navajbai Ratan Tata Trust. BCJ acknowledge support for this work from DST-SERB grant EMR/2015/000515.

\end{document}